\input harvmac.tex
\input epsf.tex
\newcount\figno
\figno=0
\def\fig#1#2#3{
\par\begingroup\parindent=0pt\leftskip=1cm\rightskip=1cm\parindent=0pt
\global\advance\figno by 1
\midinsert
\epsfxsize=#3
\centerline{\epsfbox{#2}}
\vskip 12pt
{\bf Fig. \the\figno:} #1\par
\endinsert\endgroup\par
}
\def\figlabel#1{\xdef#1{\the\figno}}
\def\encadremath#1{\vbox{\hrule\hbox{\vrule\kern8pt\vbox{\kern8pt
\hbox{$\displaystyle #1$}\kern8pt}
\kern8pt\vrule}\hrule}}

\overfullrule=0pt



\def\unlockat{\catcode`\@=11}
\def\lockat{\catcode`\@=12}

\unlockat

\def\newsec#1{\global\advance\secno by1\message{(\the\secno. #1)}
\global\subsecno=0\global\subsubsecno=0\eqnres@t\noindent
{\bf\the\secno. #1}
\writetoca{{\secsym} {#1}}\par\nobreak\medskip\nobreak}
\global\newcount\subsecno \global\subsecno=0
\def\subsec#1{\global\advance\subsecno
by1\message{(\secsym\the\subsecno. #1)}
\ifnum\lastpenalty>9000\else\bigbreak\fi\global\subsubsecno=0
\noindent{\it\secsym\the\subsecno. #1}
\writetoca{\string\quad {\secsym\the\subsecno.} {#1}}
\par\nobreak\medskip\nobreak}
\global\newcount\subsubsecno \global\subsubsecno=0
\def\subsubsec#1{\global\advance\subsubsecno by1
\message{(\secsym\the\subsecno.\the\subsubsecno. #1)}
\ifnum\lastpenalty>9000\else\bigbreak\fi
\noindent\quad{\secsym\the\subsecno.\the\subsubsecno.}{#1}
\writetoca{\string\qquad{\secsym\the\subsecno.\the\subsubsecno.}{#1}}
\par\nobreak\medskip\nobreak}

\def\subsubseclab#1{\DefWarn#1\xdef
#1{\noexpand\hyperref{}{subsubsection}%
{\secsym\the\subsecno.\the\subsubsecno}%
{\secsym\the\subsecno.\the\subsubsecno}}%
\writedef{#1\leftbracket#1}\wrlabeL{#1=#1}}
\lockat

\def\IL{\relax{\rm I\kern-.18em L}}
\def\IH{\relax{\rm I\kern-.18em H}}
\def\IR{\relax{\rm I\kern-.18em R}}
\def\IC{\relax\hbox{$\inbar\kern-.3em{\rm C}$}}
\def\IT{\relax\hbox{$\inbar\kern-.3em{\rm T}$}}
\def\IZ{\relax\ifmmode\mathchoice
{\hbox{\cmss Z\kern-.4em Z}}{\hbox{\cmss Z\kern-.4em Z}}
{\lower.9pt\hbox{\cmsss Z\kern-.4em Z}}
{\lower1.2pt\hbox{\cmsss Z\kern-.4em Z}}\else{\cmss Z\kern-.4em
Z}\fi}

\def\CN {{\cal N}}

\def\CD {{\cal D}}

\def\CL {{\cal L}}


\def\CN {{\cal N}}


\def\IZ{\relax\ifmmode\mathchoice
{\hbox{\cmss Z\kern-.4em Z}}{\hbox{\cmss Z\kern-.4em Z}}
{\lower.9pt\hbox{\cmsss Z\kern-.4em Z}}
{\lower1.2pt\hbox{\cmsss Z\kern-.4em Z}}\else{\cmss Z\kern-.4em
Z}\fi}
\def\half {{1\over 2}}

\def\p{\partial}

\def\CN {{\cal N}}


\def\IZ{\relax\ifmmode\mathchoice
{\hbox{\cmss Z\kern-.4em Z}}{\hbox{\cmss Z\kern-.4em Z}}
{\lower.9pt\hbox{\cmsss Z\kern-.4em Z}}
{\lower1.2pt\hbox{\cmsss Z\kern-.4em Z}}\else{\cmss Z\kern-.4em
Z}\fi}
\def\IB{\relax{\rm I\kern-.18em B}}
\def\IC{{\relax\hbox{$\inbar\kern-.3em{\rm C}$}}}
\def\ID{\relax{\rm I\kern-.18em D}}
\def\IE{\relax{\rm I\kern-.18em E}}
\def\IF{\relax{\rm I\kern-.18em F}}
\def\IG{\relax\hbox{$\inbar\kern-.3em{\rm G}$}}
\def\IGa{\relax\hbox{${\rm I}\kern-.18em\Gamma$}}
\def\IH{\relax{\rm I\kern-.18em H}}
\def\II{\relax{\rm I\kern-.18em I}}
\def\IK{\relax{\rm I\kern-.18em K}}
\def\IP{\relax{\rm I\kern-.18em P}}
\def\IQ{\relax\hbox{$\inbar\kern-.3em{\rm Q}$}}

\def\inbar{\,\vrule height1.5ex width.4pt depth0pt}

\def\p{\partial}

\font\cmss=cmss10 \font\cmsss=cmss10 at 7pt
\def\IR{\relax{\rm I\kern-.18em R}}


\def\boxit#1{\vbox{\hrule\hbox{\vrule\kern8pt
\vbox{\hbox{\kern8pt}\hbox{\vbox{#1}}\hbox{\kern8pt}}
\kern8pt\vrule}\hrule}}
\def\mathboxit#1{\vbox{\hrule\hbox{\vrule\kern8pt\vbox{\kern8pt
\hbox{$\displaystyle #1$}\kern8pt}\kern8pt\vrule}\hrule}}


\def\inbar{\,\vrule height1.5ex width.4pt depth0pt}

\def\p{\partial}

\font\cmss=cmss10 \font\cmsss=cmss10 at 7pt
\def\IR{\relax{\rm I\kern-.18em R}}

\def\d{{\delta}}
\def\e{{\epsilon}}
\def\g{{\gamma}}

\def\la{{\lambda}}
\def\o{{\omega}}

\def\s{{\sigma}}

\def\G{{\Gamma}}



\lref\amv{M.~Atiyah, J.~Maldacena and C.~Vafa, ``An M-theory flop as a
large N duality'', hep-th/0011256.}

\lref\ss{A.~Salam and E.~Sezgin, ``d=8 Supergravity'', Nucl.\ Phys.\ B {\bf
258} (1985) 284.}

\lref\mni{J.~M.~Maldacena and C.~N\'u\~nez, ``Towards the large N
limit of pure $\CN = 1$ super Yang Mills'', Phys.\ Rev.\ Lett.\
{\bf 86} (2001) 588, hep-th/0008001.}

\lref\en{J.D.~Edelstein and C.~N\'u\~nez, ``D6 branes and M-theory
geometrical transitions from gauged  supergravity'', JHEP {\bf
0104}, (2001) 028, {\rm hep-th/0103167}.}

\lref\epr{J.~D.~Edelstein, A.~Paredes and A.~V.~Ramallo, ``Wrapped
branes with fluxes in 8d gauged supergravity'', {\rm
hep-th/0207127}.}

\lref\mn{J.~M.~Maldacena and C.~N\'u\~nez, ``Supergravity
description of field theories on curved manifolds and a no go
theorem'', {\rm hep-th/0007018}.}

\lref\CGLPi{M.~Cvetic, G.~W.~Gibbons, H.~Lu and C.~N.~Pope,
``M-theory conifolds'', Phys.\ Rev.\ Lett.\  {\bf 88} (2002)
121602, {\rm hep-th/0112098}.}

\lref\CGLPii{M. Cvetic, G. W. Gibbons, H. Lu and C. N. Pope, ``A
$G_2$ unification of the deformed and resolved conifolds", Phys.\
Lett.\ B {\bf 534} (2002) 172, {\rm hep-th/0112138}.}

\lref\ks{I.~R.~Klebanov and M.~J.~Strassler, ``Supergravity and a
confining gauge theory: Duality cascades and (chi)SB-resolution of
naked singularities'', JHEP{\bf 0008} (2000) 052, hep-th/0007191.}

\lref\cglp{M.~Cvetic, G.~W.~Gibbons, H.~Lu and C.~N.~Pope, ``New complete
non-compact Spin(7) manifolds'', hep-th/0103155.}

\lref\bst{H.~J.~Boonstra, K.~Skenderis and P.~K.~Townsend, ``The domain
wall/QFT correspondence'', JHEP{\bf 9901} (1999) 003 [hep-th/9807137].}

\lref\bvs{M.~Bershadsky, C.~Vafa and V.~Sadov, ``D-Branes and
Topological Field Theories'', Nucl.\ Phys.\ B {\bf 463} (1996)
420, hep-th/9511222.}

\lref\bjsv{M.~Bershadsky, A.~Johansen, V.~Sadov and C.~Vafa, ``Topological
reduction of 4-d SYM to 2-d sigma models'', Nucl.\ Phys.\ B {\bf 448} (1995)
166 [hep-th/9501096].}

\lref\ens{J.~D.~Edelstein, C.~N\'u\~nez and F.~A.~Schaposnik,
``Bogomol'nyi Bounds and Killing Spinors in d=3 Supergravity'',
Phys.\ Lett.\ B {\bf 375} (1996) 163, hep-th/9512117.}

\lref\imsy{N.~Itzhaki, J.~M.~Maldacena, J.~Sonnenschein and
S.~Yankielowicz, ``Supergravity and the large N limit of theories
with sixteen supercharges'', Phys.\ Rev.\ D {\bf 58} (1998)
046004, hep-th/9802042.}

\lref\vafa{C.~Vafa, ``Superstrings and topological strings at large N'',
hep-th/0008142.}

\lref\civ{F.~Cachazo, K.~Intriligator and C.~Vafa, ``A Large N
Duality via a Geometric Transition'', hep-th/0103067.}

\lref\ed{J.~D.~Edelstein, ``Large N dualities from wrapped
D-branes'', hep-th/0211204.}

\lref\eprii{J.~D.~Edelstein, A.~Paredes and A.~V.~Ramallo, ``Let's
twist again: General metrics of G(2) holonomy from gauged
supergravity'', hep-th/0211203.}

\lref\jaume{J.~Gomis, ``D-Branes, Holonomy and M-Theory'',
hep-th/0103115.}

\lref\cdlo{P.~Candelas and X.~C.~de la Ossa, ``Comments On Conifolds'',
Nucl.\ Phys.\ B {\bf 342} (1990) 246.}

\lref\pzt{L.~A.~Pando Zayas and A.~A.~Tseytlin, ``3-branes on
resolved conifold'', JHEP{\bf 0011} (2000) 028, hep-th/0010088.}

\lref\hsi{R.~Hernandez and K.~Sfetsos, ``An eight-dimensional
approach to G(2) manifolds'', Phys.\ Lett.\ B {\bf 536} (2002)
294, {\rm hep-th/0202135}.}

\lref\bs{R.~Bryant and S.~Salamon, ``On the construction of some complete
metrics with exceptional holonomy'', Duke\ Math.\ J.\ {\bf 58} (1989) 829.}

\lref\gpp{G.~W.~Gibbons, D.~N.~Page and C.~N.~Pope, ``Einstein Metrics On
$S^3$, $R^3$ And $R^4$ Bundles'', Commun.\ Math.\ Phys.\ {\bf 127} (1990)
529.}

\lref\pt{G.~Papadopoulos and A.~A.~Tseytlin, ``Complex geometry of
conifolds and 5-brane wrapped on 2-sphere'', Class.\ Quant.\
Grav.\  {\bf 18}, 1333 (2001), hep-th/0012034.}

\lref\pztdos{L.~A.~Pando Zayas and A.~A.~Tseytlin, ``3-branes on
spaces with $\IR\times S^2\times S^3$ topology'', Phys. Rev. D{\bf 63} (2001)
086006, hep-th/0101043.}

\lref\hsiii{R.~Hernandez and K.~Sfetsos, ``Holonomy from wrapped branes'',
hep-th/0211130.}

\lref\luvp{H.~L\"u and J.~F.~Vazquez-Poritz, ``$S^1$-wrapped D3-branes on
conifolds,'' Nucl.\ Phys.\ B {\bf 633}, 114 (2002), hep-th/0202175.}

\Title{\vbox{\baselineskip12pt \hbox{US-FT-6/02}
\hbox{hep-th/0212139} }} {\vbox{\centerline{Singularity Resolution
in Gauged Supergravity}
\medskip
\centerline{and Conifold Unification}}} \centerline{J. D.
Edelstein${}^{\,\sharp\,\dagger\,*}$
\foot{jedels@math.ist.utl.pt}, A. Paredes${}^{\,*}$
\foot{angel@fpaxp1.usc.es} and A.V. Ramallo${}^{\,*}$
\foot{alfonso@fpaxp1.usc.es}}

\bigskip
\medskip
{\vbox{\centerline{${}^{\sharp\,}$ \sl Departamento de
Matem\'atica, Instituto Superior Tecnico} \centerline{\sl Av.
Rovisco Pais, 1049--001, Lisboa, Portugal}}}
\medskip
{\vbox{\centerline{${}^{\dagger\,}$ \sl Instituto de F\'\i sica de
La Plata -- Conicet, Universidad Nacional de La Plata}
\centerline{\sl C.C. 67, (1900) La Plata, Argentina}}}
\medskip
{\vbox{\centerline{${}^{\,*}$Departamento de F\'\i sica de
Part\'\i culas, Universidad de Santiago de Compostela}
\centerline{\sl E-15782 Santiago de Compostela, Spain}}}

\bigskip
\bigskip
\noindent We obtain a unified picture for the conifold singularity
resolution. We propose that gauged supergravity, through a novel
prescription for the twisting, provides an appropriate framework
to smooth out singularities in the context of gravity duals of
supersymmetric gauge theories.

\Date{11 December 2002}


\newsec{Introduction}

String theory compactifications in a Calabi--Yau threefold have
been the focus of a countless number of papers due to the fact
that they provide effective four dimensional vacua with ${\cal
N}=1$ supersymmetry. Particular emphasis has been given to the
study of singularities on these manifolds --in particular, conical
singularities-- as long as non-trivial phenomena take place on
them such as gauge symmetry enhancement or the appearance of new
massless particles. The archetype of these is the well--known
conifold. It is a complex three--manifold which is a cone over the
homogeneous space $T^{1,1} = {SU(2) \times SU(2) \over U(1)}$. The
conical singularity can be resolved in two different ways
according to whether an $S^2$ or an $S^3$ is blown up at the
singular point. The former is known as the {\it resolved} (or
K\"ahler deformed) conifold, while the latter is the (complex)
{\it deformed} conifold. Both regular manifolds depend on a single
parameter (namely, the {\it resolution} $a$ and the {\it deformation}
$\mu$), are non compact and asymptotically behave as the singular
conifold. That is, the three solutions display the same UV
behavior for the associated gauge theories. Supersymmetry and
matching holonomy conditions in the context of String theory
\jaume\ ensure that there must exist manifolds with $G_2$ holonomy
metrics whose Gromov--Hausdorff limits are precisely the Ricci
flat K\"ahler metrics on the resolved and deformed conifolds.
These were explicitly found in \CGLPi. It was shown afterwards
that these $G_2$ manifolds arise as solutions of the same system
of first--order equations, this providing a nicely unified picture
of the resolved and deformed conifolds from the perspective of
M--theory \CGLPii.

In a different approach based on lower dimensional gauged
supergravity, it was recently shown that the resolved conifold
comes out when studying the gravity dual of $D6$--branes wrapping
an holomorphic $S^2$ in a $K3$ manifold \en. The low--energy
dynamics is governed, when the size of the cycle is taken to zero,
by a five dimensional supersymmetric gauge theory with eight
supercharges. If the manifold is large enough and smooth, the dual
description is given in terms of a purely gravitational
configuration of eleven dimensional supergravity which is the
direct product of Minkowski five dimensional spacetime and the
resolved conifold. The general solution to this system was later
shown \epr\ to be given by the {\it generalized} resolved conifold
\cdlo\pt\pzt\pztdos. This is a one-parameter (say, $b$) generalization
of the resolved conifold. There is an analogous extension metric both
for the deformed conifold --though it is not regular--, and the singular
conifold. We will call the latter {\it regularized} conifold,
following \pztdos, because $b$ smoothens the curvature singularity
and the metric is regular upon imposing a $\IZ_2$ identification
of the $U(1)$ fiber. It is an ALE space that asymptotically
approaches $T^{1,1}/\IZ_2$.

Lower dimensional gauged supergravities provide an explicit arena
to impose the twisting conditions required to wrap a $D$--brane in
a supersymmetric cycle \bvs\mn. (See \ed\ for a recent review.)
Loosely speaking, in the conventional twist, the gauge connection has
to be identified with the spin connection. This notion of the twist can
be generalized, as shown in \hsi, in a way that involves non-trivially
the scalar fields that arise in lower dimensional gauged supergravity
from the external components of the metric. The solutions obtained by
these means usually correspond to the near horizon limit of wrapped
D--branes \bst. However, in a recent paper \eprii, we have shown that
the twist can be further generalized so that it encompasses much more
general solutions either corresponding to wrapped D--branes or to
special holonomy manifolds with certain RR fluxes turned on. On the
one hand, the new twisting condition can be thought of as a
non-trivial embedding of the world--volume in the lower dimensional
spacetime. More interestingly, as we will show in this paper through
an archetypical example, it can also be understood as a singularity
resolution mechanism \foot{Notice that, in a sense, it is natural to
expect that lower dimensional gauged supergravity degrees of freedom
cure singularities. For example, even when using the standard twisting
conditions, the resolution of the conifold singularity has been shown
to be driven by turning on a scalar field in gauged supergravity \en.}:
the ordinary twisting imposes the value of the gauge fields at infinity,
while the lower dimensional gauged supergravity governs the nontrivial
dynamics in the bulk. This mechanism resembles that used in the
Maldacena--N\'u\~nez solution \mni, where the singularity is solved
by turning on a non--Abelian gauge field that asymptotically approaches
the Abelian one that twists the gauge theory.

In this letter we present a unified scenario for
conifold singularity resolutions. In a sense, we are providing the
unified picture of the resolved and deformed conifolds from the
perspective of M--theory advocated in \CGLPii. The main difference
being that we deal with conifolds in eleven dimensions instead of
$G_2$ manifolds. A unique system encompass the generalized
resolution and deformation of the conifold singularity, each of
them emerging as the only two possible solutions of an algebraic
constraint. Notice the difference with the $G_2$ case studied in
\eprii, where the algebraic constraints are involved enough so as
to admit several well distinct solutions. Here there are only two.
Furthermore, we show that it is possible to impose at the same time
both solutions of the algebraic constraint, this leading to the
regularized conifold metric, which describes a complex line bundle
over $S^2 \times S^2$.

\newsec{D6--branes wrapped on ${\bf S^2}$ revisited}

The Lagrangian describing the dynamics of the sector of Salam and
Sezgin's eight dimensional gauged supergravity \ss\ on which we
would like to focus (entirely coming from the eleven dimensional
metric), reads 
\eqn\boslag{
e^{-1} \CL = {1 \over 4} R - {1 \over
4} e^{2\phi} (F_{\mu\nu}^{~i})^2 - {1 \over 4} (P_{\mu ij})^2 -
\half (\p_\mu\phi)^2 - {1 \over 32} e^{-2\phi} (e^{-8\la} - 4
e^{-2\la}) ~,} 
where $\phi$ is the dilaton, $\la$ is a scalar in
the coset $SL(3,\IR)/SO(3)$, $A^i$ is an $SU(2)$ gauge potential,
$e$ is the determinant of the vierbein $e^a$, $F^i$ is the
Yang--Mills field strength and $P_{ij}$ is a symmetric and
traceless 1--form defined by 
\eqn\cincuenta{
P_{ij} + Q_{ij} =
\pmatrix{d\lambda&&-A^{3}&&A^{2}\,e^{-3\lambda}\cr\cr
A^{3}&&d\lambda&&-A^{1}\,e^{-3\lambda}\cr\cr -A^{2}
e^{3\lambda}&&A^{1}\,e^{3\lambda}&&-2d\lambda} ~,} 
$Q_{ij}$ being
the antisymmetric counterpart. The supersymmetry transformations
for the fermions are given by 
\eqn\stpsi{
\d\psi_\g = \CD_\g \e +
{1 \over 24} e^{\phi} F_{\mu\nu}^i \hat\G_i (\G_\g^{~\mu\nu} - 10
\d_\g^{~\mu} \G^\nu) \e - {1 \over 288} e^{-\phi} \e_{ijk}
\hat\G^{ijk} \G_\g T \e ~, ~~~~~~~} 
\eqn\stchi{
\d\chi_i = \half
(P_{\mu ij} + {2 \over 3} \d_{ij} \p_\mu\phi) \hat\G^j \G^\mu \e -
{1 \over 4} e^{\phi} F_{\mu\nu i} \G^{\mu\nu} \e - {1 \over 8}
e^{-\phi} (T_{ij} - \half \d_{ij} T) \e^{jkl} \hat\G_{kl} \e ~,}
where $T_{ij} = {\rm diag}(e^{2\la},e^{2\la},e^{-4\la})$, $T =
\delta^{ij} T_{ij} = 2 e^{2\la} + e^{-4\la}$, and the covariant
derivative is 
\eqn\covd{ 
\CD \e ~=~ \bigl( \p + {1 \over 4} \o^{ab} \G_{ab} + {1 \over 4}
Q_{ij} \hat\G^{ij} \bigr) \e ~.} 
We shall use the following representation for the Dirac
matrices: 
\eqn\dirac{
\Gamma^{ \mu}\,=\,\gamma^{\mu} \otimes \II ~,
~~~~~ \,\,\,\,\, \hat\Gamma^{ i}\,=\,\gamma_9\,\otimes\,\tau^i ~,}
where $\gamma^{\mu}$ are eight dimensional Dirac matrices,
$\gamma_9\,=\,i \gamma^{ 0}\, \gamma^{ 1}\,\cdots\,\gamma^{7}$
($\gamma_9^2\,=\,1$), $\tau^i$ are Pauli matrices and
$\hat\Gamma^{ i}$ are the Dirac matrices along the $SU(2)$ group
manifold, whereas $\Gamma_7\equiv\Gamma_r$ corresponds to the
radial direction.

We shall consider the following ansatz for the eight dimensional
metric 
\eqn\metr{
ds^2_8 = e^{2f} dx_{1,4}^2 + e^{2h} d\Omega_2^2 +
dr^2 ~,} 
where $d\Omega_2^2 = d\theta^2 + \sin^2\theta d\varphi^2$
is the metric of the unit $S^2$. The ansatz for the gauge field is
better presented in terms of the triplet of Maurer--Cartan 1-forms
on $S^2$ 
\eqn\maucar{\s^1 = d\theta ~, ~~~~~ \s^2 = \sin\theta
d\varphi ~, ~~~~~ \s^3 = \cos\theta d\varphi ~,} 
that obey the
conditions $d\s^i = - \ha \e_{ijk} \s^j\wedge \s^j$. It is:
\eqn\connec{
A^{1} = g(r) ~\s^1 ~, ~~~~~ \,\,\, A^{2} = g(r) ~\s^2
~, ~~~~~ \,\,\, A^{3} = \s^3 ~.} 
Notice that the twisting in \en\
corresponds to $g(r) = 0$. We will check at the end that $g(r) \to
0$ asymptotically so, from the point of view of the dual gauge
theory, the twisting is not modified. The field strength, $F^i =
dA^i + {1\over 2}\,\epsilon_{ijk}\,A^j\wedge A^k$, reads:
\eqn\fstrength{
F^{1} = g' ~dr \wedge \s^1 ~, ~~~~~ \,\,\, F^{2} =
g' ~dr \wedge \s^2 ~, ~~~~~ \,\,\, F^{3} = \big(g^2-1) ~\s^1 \wedge
\s^2 ~.} 
When uplifted to eleven dimensions, the unwrapped part of
the metric should correspond to flat five dimensional Minkowski
spacetime. This condition determines the relation $f=\phi/3$ that
we impose from now on. Actually, it is not difficult to write down the
form of the eleven dimensional metric for the ansatz we are adopting. 
Let $ w^i$ for $i=1,2,3$ be a set of SU(2) left invariant one forms of
the external three sphere satisfying $dw^i =  \ha \e_{ijk} w^j\wedge w^k$. 
Then, the uplifted eleven dimensional metric is: 
\eqn\upliftmet{
\eqalign{ds^2_{11} = & dx_{1,4}^2\,+\,
e^{2h-{2\phi\over 3}}\,d\Omega_2^2\,+\,e^{-{2\phi\over
3}}dr^2\,+\, 4\,e^{{4\phi\over 3}+2\lambda}\,\big(\,
w^1+\,g\sigma^1\,\big)^2 \,+\,\cr & + \,4\,e^{{4\phi\over
3}+2\lambda}\, \big(\, w^2+\,g\sigma^2
\,\big)^2\,+\, 4\,e^{{4\phi\over 3}-4\lambda}
\big(\, w^3+\,\sigma^3\,\big)^2 ~.}} 

In order to seek for supersymmetric solutions to the system, we
start by subjecting the spinor to the following angular projection
\eqn\ktproj{ 
\Gamma_{\theta\varphi} \epsilon = -\hat\Gamma_{12}
\epsilon ~,} 
which is imposed by the K\"ahler structure of the
ambient $K3$ manifold in which the two--cycle lives. The equations
$\delta\chi_1\,=\delta\chi_2=0$ give: 
\eqn\dchii{
\Big(\lambda' +
{2\over 3} \phi' \Big) \epsilon = g e^{-h} \sinh 3\lambda\,\,
\hat\Gamma_1 \Gamma_{\theta} \Gamma_{r} \hat\Gamma_{123} \epsilon
- e^{\phi+\lambda-h} g' \hat\Gamma_1\Gamma_{\theta} \epsilon  -
{1\over 4} e^{-\phi-4\lambda}\,\Gamma_{r} \hat\Gamma_{123}\epsilon
~,} 
while $\delta\chi_3\,=0$ reads: 
\eqn\dchit{
\eqalign{
\Big(\,2\lambda'\,-\,{2\over 3}\,\phi'\,\Big)\,\epsilon = &
\Big[\,e^{\phi-2\lambda-2h}\,\big(g^2-1)\,-\, {1\over
4}\,e^{-\phi}\,\big(\,e^{-4\lambda}\,-\,2e^{2\lambda}\,\big)\,
\Big]\,\Gamma_{r}\,\hat\Gamma_{123}\,\epsilon \cr & +
2g\,e^{-h}\,\sinh 3\lambda\,\,\hat\Gamma_1\,\Gamma_{\theta}\,
\Gamma_{r}\,\hat\Gamma_{123}\,\epsilon ~.}} 
One can combine these
two equations to eliminate $\lambda'$: 
\eqn\phipr{
\phi'\epsilon +
e^{\phi+\lambda-h} g' \hat\Gamma_1\Gamma_{\theta} \epsilon
\,+\,\Big[\,{1\over 2}\,e^{\phi-2\lambda-2h} \big(g^2-1)\,+\,
{1\over 8}\,e^{-\phi}\,\big(\,e^{-4\lambda}\,+\,2e^{2\lambda}
\big)\, \Big]\,\Gamma_{r}\,\hat\Gamma_{123}\,\epsilon = 0 ~.} 
From
this last equation, it is clear that the supersymmetric parameter
must satisfy a projection of the sort: 
\eqn\projt{
\Gamma_{r}\,\hat\Gamma_{123}\,\epsilon\,=\,-\big(\, \beta +
\tilde\beta\,\hat\Gamma_1\,\Gamma_{\theta}\,\big)\,\epsilon ~,}
where $\beta$ and $\tilde\beta$ are (functions of the radial
coordinate) proportional to the first derivatives of $\phi'$ and
$g'$: 
\eqn\phpr{
\phi' = \Big[\, {1\over
2}\,e^{\phi-2\lambda-2h}\,\big(g^2-1)\,+\, {1\over 8} e^{-\phi}
\big(\,e^{-4\lambda}\,+\,2e^{2\lambda}\,\big)\, \Big]\,\beta ~,}
\eqn\gprim{
e^{\phi+\lambda-h}\,g' = \Big[\, {1\over 2}
e^{\phi-2\lambda-2h} \big(g^2-1)\,+\, {1\over 8} e^{-\phi} \big(
e^{-4\lambda} +\,2e^{2\lambda}\,\big)\, \Big]\,\tilde\beta ~.}
This radial projection provides the generalized twist introduced in
\eprii. It encodes a non-trivial fibering of the two sphere with the
external three sphere as will become clear below. Since $(\Gamma_r\,
\hat\Gamma_{123})^2\epsilon\,=\,\epsilon$ and $\{\Gamma_{r}
\hat\Gamma_{123}, \hat\Gamma_1 \Gamma_{\theta}\}=0$, one must have
$\beta^2\,+\,\tilde\beta^2 = 1$. Thus, we can
represent $\beta$ and $\tilde\beta$ as 
\eqn\betapar{
\beta\,=\,\cos\alpha ~, ~~~~~\,\, \tilde\beta\,=\,\sin\alpha ~.}
Also, it is clear that both independent projections \ktproj\ and
\projt\ leave unbroken eight supercharges as expected.
Inserting the radial projection \projt, as well as \phpr, in
\dchit, we get an equation determining $\lambda'$: 
\eqn\lapr{
\lambda' =\,ge^{-h}\,\sinh 3\lambda\,\,\tilde\beta\,-\,
\Big[\,{1\over 3}\,e^{\phi-2\lambda-2h}\,\big(g^2-1)\,-\, {1\over
6}\,e^{-\phi}\,\big(\,e^{-4\lambda}\,-\,e^{2\lambda}\,\big)\,
\Big]\,\beta ~,} 
together with an algebraic constraint:
\eqn\const{
ge^{-h}\,\sinh 3\lambda\,\,\beta\,+\, \Big[\,{1\over
2}\,e^{\phi-2\lambda-2h}\,\big(g^2-1)\,-\, {1\over
8}\,e^{-\phi}\,\big(\,e^{-4\lambda}\,-\,2e^{2\lambda}\,\big)\,
\Big]\,\tilde\beta\,=\,0 ~.} 
Let us now consider the equations
obtained from the supersymmetric variation of the gravitino. From
the components along the unwrapped directions one does not get
anything new, while from the angular components we get:
\eqn\gravit{
\eqalign{h'\epsilon = & -\,ge^{-h}\,\cosh 3\lambda
\hat\Gamma_1 \Gamma_{\theta}\, \Gamma_{r}\,\hat\Gamma_{123}
\epsilon + {2\over 3} e^{\phi+\lambda-h} g' \hat\Gamma_1
\Gamma_{\theta}\,\epsilon \cr & - {1\over 6}\Big[-5
e^{\phi-2\lambda-2h}\,\big(g^2-1)\,+\, {1\over 4}\,e^{-\phi}
\big(\,2e^{2\lambda}\,+\,e^{-4\lambda}\,\big)\, \Big]\,\Gamma_{r}
\hat\Gamma_{123}\,\epsilon ~.}}
By using the projection \projt\ we obtain an equation for $h'$: 
\eqn\hprim{
h'\,=\,-ge^{-h}\,\cosh
3\lambda\, \tilde\beta +\, {1\over 6}\Big[\,-5
e^{\phi-2\lambda-2h} \big(g^2-1)\,+\, {1\over 4}\,e^{-\phi}
\big(\,2e^{2\lambda} + e^{-4\lambda} \big)\, \Big]\,\beta ~,}
together with a second algebraic constraint:
\eqn\algcons{
-ge^{-h}\,\cosh 3\lambda\, \beta +\, \Big[\,{1\over
2}\,e^{\phi-2\lambda-2h} \big(g^2-1)\,-\, {1\over
8}\,e^{-\phi}\,\big(\,2e^{2\lambda} + e^{-4\lambda} \big)\,
\Big]\,\tilde\beta\,=\,0 ~.} 
Finally, from the radial component of the gravitino we get the
functional dependence of the supersymmetric parameter $\epsilon$: 
\eqn\radgrav{
\partial_r\,\epsilon = {5\over
6}\,e^{\phi+\lambda-h}\,g'\,\, \hat\Gamma_1\,\Gamma_{\theta}
\epsilon - {1\over 12}\, \Big[\,e^{\phi-2\lambda-2h} \big(g^2-1) +
{1\over 4}\, \big(\,2e^{2\lambda} + e^{-4\lambda}\,\big)\,\Big]\,
\Gamma_{r}\,\hat\Gamma_{123}\,\epsilon ~.}

The projection \projt\ gives the generalized twisting conditions
first studied in \eprii. Its interpretation follows from a similar
reasoning as the one used in that reference: using the trigonometric
parametrization \betapar, the
generalized projection can be written as: 
\eqn\newproj{
\Gamma_{r}\,\hat\Gamma_{123} ~\epsilon = - e^{\alpha\hat\Gamma_1
\Gamma_{\theta}} ~\epsilon ~,} 
which can be solved as:
\eqn\solvpr{
\epsilon\,=\,e^{-{1\over 2}\alpha\hat\Gamma_1 \Gamma_{\theta}}\,
\epsilon_0 ~, ~~~~~~~~ \Gamma_{r} \hat\Gamma_{123}
~\epsilon_0 = -\epsilon_0 ~.}
We can determine $\epsilon$ by plugging \solvpr\ into the equation for
the radial component of the gravitino \radgrav. Using \newproj, we get
two equations. The first one gives the characteristic radial dependence
of $\epsilon_0$ in terms of the eight dimensional dilaton, namely:
\eqn\radep{
\partial_r \epsilon_0\,= {\phi'\over 6}\,\epsilon_0 ~
\,\,\,\,\,\,\,\Rightarrow \,\,\,\,\,\,\,
\epsilon_0\,= e^{{\phi\over 6}}\,\eta ~,} 
with $\eta$
being a constant spinor. The other equation determines the radial
dependence of the phase $\alpha$: 
\eqn\radalpha{\alpha' =
-2e^{\phi+\lambda-h} g' ~.} Thus, the spinor $\epsilon$ can be
written as: 
\eqn\spinor{ 
\epsilon\,=\,e^{{\phi\over 6}}\,
e^{-{1\over 2}\alpha\hat\Gamma_1 \Gamma_{\theta}}\,\eta\,\,,
~~~~~ \,\,\,\, \Gamma_{r}\,\hat\Gamma_{123}\,\eta\,=\,-\eta\,\,,
~~~~~ \,\,\,\,\,\, \Gamma_{\theta\varphi}\,\hat\Gamma_{12}\,\eta =
\eta ~.} 
The meaning of the phase $\alpha$ can be better
understood by using the following $\Gamma$--matrices identity
$\Gamma_{x^0\cdots x^4} \Gamma_{\theta\varphi} \Gamma_{r}
\hat\Gamma_{123} \,=\,-1$, so that
\eqn\whered{\Gamma_{x^0\cdots
x^4}\,\big(\,\cos\alpha ~ \Gamma_{\theta\varphi} - \sin\alpha
~\Gamma_{\theta}\hat\Gamma_{2} \big) \epsilon =\,\epsilon ~,}
which shows that the D6--brane is wrapping a two--cycle which is
non-trivially embedded in the $K3$ manifold as seen from the
uplifted perspective that is implied in \whered. Indeed, the case
$\alpha = 0$ corresponds to the D6--brane wrapping a two--sphere
that is fully contained in the eight dimensional spacetime where
supergravity lives, studied in \en. 

\newsec{Solution of the algebraic constraints}

In the previous section we derived two algebraic constraints
\const\ and \algcons\ that the functions of our ansatz must obey.
Let us presently solve them. By adding and subtracting the two
equations, we get: 
\eqn\solvi{
\tan\alpha \equiv {\tilde \beta\over
\beta}\,=\,-2\,g e^{\phi+\lambda-h}\,=\,{g\,e^{-3\lambda-h}\over
e^{\phi-2\lambda-2h} \big(g^2-1)\,-\,{1\over 4}\,
e^{-\phi-4\lambda}} ~.} 
Whereas the first part of this equation allows us to write
$\alpha$ in terms of the remaining functions, the
last equality provides an algebraic constraint that restricts our
ansatz. It is not hard to arrive at the following simple equation:
\eqn\constraint{
g\,\Big[\,g^2-1\,+\,\,{1\over 4}\,e^{-2\phi-2\lambda+2h} 
\, \Big]\,=\,0 ~.} 
There are obviously two solutions. The first one is clearly $g=0$,
which corresponds to $\tilde\beta=0$, $\beta=1$, or $\alpha=0$.
One can check that this is a consistent truncation of the system
of equations that actually reduce to the case studied in \en,
whose integral is the generalized resolved conifold (see also \epr).
Indeed, the resulting eleven dimensional metric can be written as
$ ds^2_{11} = dx_{1,4}^2 + ds^2_{6}$, where the six dimensional metric 
$ds^2_{6}$ is:
\eqn\resolved{
\eqalign{
ds^2_6\,&=\,\big[\,\kappa(\rho)\,\big]^{-1}\,d\rho^2\,+\,
{\rho^2\over 9}\,\,\kappa(\rho)\,
(\,d\psi\,+\sum_{a=1,2}\,\cos\theta_a\,d\phi_a\,)^2\cr
&+\,{1\over 6}\,(\,\rho^2\,+\,6a^2\,)\,
(\,d\theta_1^2\,+\,\sin^2\theta_1\,d\phi_1^2\,)\,+\,
{1\over 6}\,\rho^2\, (\,d\theta_2^2\,+\,\sin^2\theta_2\,d\phi_2^2\,)\,\,,}
}
with $\kappa(\rho)$ being:
\eqn\k{
\kappa(\rho)\,=\,{\rho^6\,+\,9a^2\,\rho^4\,-\,b^6\over
\rho^6\,+\,6a^2\rho^4}\,\,.}
In eq. \resolved, $\rho$ is a new radial variable, $\theta_1\equiv\theta$, 
$\phi_1\equiv\varphi$ and $(\theta_2, \phi_2, \psi)$ parametrize the 
$w^i$'s. The constants $a$ and $b$ provide the generalized resolution
of the conifold singularity \pt\pztdos. In the context of gauged
supergravity, even when this solution corresponds to the conventional
twist, $a$ and $b$ are non-zero when certain scalar fields are excited
\en\epr. The case $a=0$, $b\not=0$ corresponds to the above
mentioned regularized conifold \pztdos.

The other solution to eq. \constraint\ gives a non trivial relation
between $g$ and the remaining functions of the ansatz, namely: 
\eqn\othersol{
g^2\,=\,1\,-\,{1\over 4}\,e^{-2\phi-2\lambda+2h} ~.} 
It is not difficult to find the values of
$\beta$ and $\tilde\beta$ for this solution of the constraint:
\eqn\betais{\beta\,=\,{1\over
2}\,e^{-\phi-\lambda+h} ~, ~~~~~ \,\,\,\,\, \tilde\beta\,=\,-g ~.}
Notice that they satisfy $\beta^2+\tilde\beta^2=1$ as a consequence of
the relation \othersol. Moreover, one can verify that \othersol\ is
consistent with the first-order equations. Indeed, by differentiating
eq. \othersol\ and using the first-order equations for $\phi$, $\lambda$
and $h$ (eqs. \phpr, \lapr\ and \hprim), we arrive precisely at the
first-order equation for $g$ written in \gprim. It can be also checked
that eq. \radalpha\ is identically satisfied for this solution of the
constraint. Thus, one can eliminate $g$ from the first-order equations
arriving at the following system of equations for $\phi$, $\lambda$ and
$h$: 
\eqn\bpseqs{
\eqalign{
\phi' = & {1\over 8}\,e^{-2\phi+\lambda+h}\,\,,\cr \lambda' = &
{1\over 24}\,e^{-2\phi+\lambda+h}\,-\, {1\over 2}\,e^{3\lambda-h}
+ {1\over 2}\,e^{-3\lambda-h}\,\,,\cr h' = & -{1\over 12}
e^{-2\phi+\lambda+h} +\,{1\over 2}\,e^{3\lambda-h} + {1\over
2}\,e^{-3\lambda-h} ~.}}

\newsec{The generalized deformed conifold}

In order to integrate the system \bpseqs,  
let us define the function $z\,=\,\phi+\lambda-h$  and a new
radial coordinate $\tau$, $dr\,=\,2\,e^{\phi-2\lambda}\,d\tau$.
Then, if the dot denotes the derivative with respect to $\tau$,
it follows from \bpseqs\ that $z$ satisfies the equation: 
\eqn\zetaeq{
\dot z =
{1\over 2} e^{-z} - 2e^{z} ~.} This equation can be immediately
integrated: 
\eqn\ezeta{
e^z\,=\,{1\over 2}\,\,{\cosh
(\tau+\tau_0)\over \sinh(\tau+\tau_0)} ~,} 
where $\tau_0$ is an integration constant, which from now on we will 
absorb in a redefinition of the origin of $\tau$. We can obtain $\phi$ by
noticing that it satisfies the equation: 
\eqn\phieq{
\dot\phi\,=\,{1\over 4}\,e^{-z}~ .} 
Since we know $z(\tau)$ explicitly, we can obtain immediately
$\phi(\tau)$, namely: 
\eqn\phisolv{
e^{\phi}\,=\,\hat\mu
\big(\cosh\tau\big)^{{1\over 2}} ~,}
where $\hat\mu$ is a constant
of integration. Finally, $h$ satisfies the following differential
equation: 
\eqn\heq{
\dot h\,=\,-{1\over 6}\,e^{-z}\,+\,e^{z}\,+
e^{6\phi-5z-6h} ~.} 
If we define, $y=e^{6h}$ and use the
expressions of $z$ and $\phi$ as functions of $\tau$, we get:
\eqn\yeq{
\dot y\,=\,{\cosh^2\tau+2\over \cosh\tau\sinh\tau}\,y +
192\,\hat\mu^6\,\,{(\sinh\tau)^5\over (\cosh\tau)^2} ~,} 
which is
also easily integrated by the method of variation of constants. In
order to express the corresponding result, let us define the
function: 
\eqn\funk{
K(\tau)\equiv {\Big(\sinh 2\tau\,-\,2\tau
+\,C\,\Big)^{1\over 3}\over 2^{{1\over 3}}\,\sinh\tau} ~,} 
where $C$ is a new constant of integration. Then, $h$ is given by:
\eqn\hsolii{
e^{h}\,=\,3^{{1\over 6}}\,2^{{5\over 6}}\,\hat\mu\,\,
{\sinh\tau\over (\cosh\tau)^{{1\over 3}}}\, \big[ K(\tau)
\big]^{{1\over 2}} ~.} 
As we know $z$, $h$ and $\phi$, we can
obtain $\lambda$. The result is: 
\eqn\lambsol{ 
e^{\lambda} = \Big(
{3\over 2} \Big)^{{1\over 6}}\, (\cosh\tau)^{{1\over 6}}\, \big[
K(\tau) \big]^{{1\over 2}} ~.} 
Finally, we can get $g$ from the
solution of the constraint (eq. \othersol), namely: 
\eqn\gsolv{
g\,=\,{1\over
\cosh\tau} ~.} 
It follows immediately from \gsolv\ that $g\rightarrow 0$ as
$\tau\rightarrow\infty$, as claimed above. Moreover,  by  using the
explicit form of this solution we can find the value of the phase $\alpha$: 
\eqn\alphase{
\cos\alpha\,=\,{\sinh\tau\over\cosh\tau} ~, ~~~~~ \,\,\,\,\,
\sin\alpha\,=\,-{1\over \cosh\tau} ~,}
Notice that $\alpha\rightarrow -\pi/2$ when $\tau\rightarrow 0$, whereas
$\alpha\rightarrow 0$ for $\tau\rightarrow\infty$. In order to express
neatly the form of the corresponding eleven dimensional metric, let us
define the following set of one-forms:
\eqn\gbein{
\eqalign{g^1 = & {1\over \sqrt{2}}\,
\big[\,\sigma^2\,-\, w^2\,\big]\,,~~~~~ \,\,\,\,\,
g^2= {1\over \sqrt{2}}\, \big[\,\sigma^1\,-\, w^1\,\big]\,,\cr
g^3 = & {1\over \sqrt{2}}\, \big[\,\sigma^2\,+\, w^2\,\big]\,,
~~~~~ \,\,\,\,\,g^4 =  {1\over \sqrt{2}}\,
\big[\,\sigma^1\,+\, w^1\,\big]\,,\cr 
g^5 = & {1\over \sqrt{2}}\, 
\big[\,\sigma^3\,+\, w^3\,\big] ~,}}
and a new constant $\mu$, related to $\hat\mu$ as $\mu =
2^{{11\over 4}} 3^{{1\over 4}} \hat\mu$. Then, by using the uplifting
formula \upliftmet, the resulting eleven
dimensional metric $ds^2_{11}$ can again be written as $ ds^2_{11} =
dx_{1,4}^2 + ds^2_{6}$, where now the six dimensional metric is:
\eqn\gendefcon{
\eqalign{ds^2_{6}={1\over 2}\,\mu^{{4\over
3}}\,K(\tau) \Bigg[&{1\over 3 K(\tau)^3}\,\big(\,d\tau^2\,+\,
(g^5)^2\,)\,+\, \cosh^2\big({\tau\over 2}\big)\,
\big(\,(g^3)^2\,+\, (g^4)^2\,)\,+\,\cr &+\,\sinh^2\big({\tau\over
2}\big)\,\big(\,(g^1)^2\,+\, (g^2)^2\,)\, \,\Bigg]\,\,,}} 
which, for $C=0$ is nothing but the standard metric of the deformed
conifold, with $\mu$ being the corresponding deformation parameter. 

The metric \gendefcon\ for $C\not=0$ was studied in ref. \pztdos, where
it was shown to display a curvature singularity when $\mu\not=0$. On the
contrary, for $\mu=0$ and $C\not=0$ this metric is regular, after a
$\IZ_2$ identification of the $U(1)$ fiber, and reduces to the one
written in \resolved\ for $a=0$ and $b\not=0$ (the regularized conifold),
the parameter $b$ being related to the constant $C$ \pztdos. It is not
difficult to reobtain this result within our formalism. Notice, first of
all, that both solutions of the constraint \constraint\ are not
incompatible, {\rm i.e.} one can take $g=0$ in eq. \othersol\ if
$z=\phi+\lambda-h$ is fixed to the particular constant value 
$e^z=1/2$. Notice that this is consistent with eq. \zetaeq. Actually, 
this value of $z$ can be obtained by taking $\tau_0\rightarrow\infty$ in
the general solution \ezeta. Moreover, the differential equation \phieq\
for $\phi$ in this $g=0$ case reduces to $\dot\phi=1/2$, which can be
immediately integrated to give $e^{\phi}=Ae^{\tau/2}$, with $A$ being a
non-zero constant. Again, this solution can be obtained from the general
expression \phisolv\ by first reintroducing the $\tau_0$ parameter ({\rm
i.e.} by changing $\tau\rightarrow\tau+\tau_0$) and then by taking 
$\tau_0\rightarrow\infty$ and $\hat\mu\rightarrow 0$ in such a way that 
$\hat\mu\,e^{\tau_0/2}=\sqrt{2}\,A$. Notice that this corresponds to taking
$\mu=0$, as claimed. It follows from this discussion that the regularized
conifold is a boundary in the moduli space separating the regions that
correspond to the generalized deformed and resolved conifolds, as depicted
in the figure. Notice that we cannot continously connect the deformed
and resolved conifolds through a supersymmetric trajectory of non-singular
metrics.

\bigskip
\centerline{\vbox{\hsize=5in\tenpoint
\centerline{\epsffile{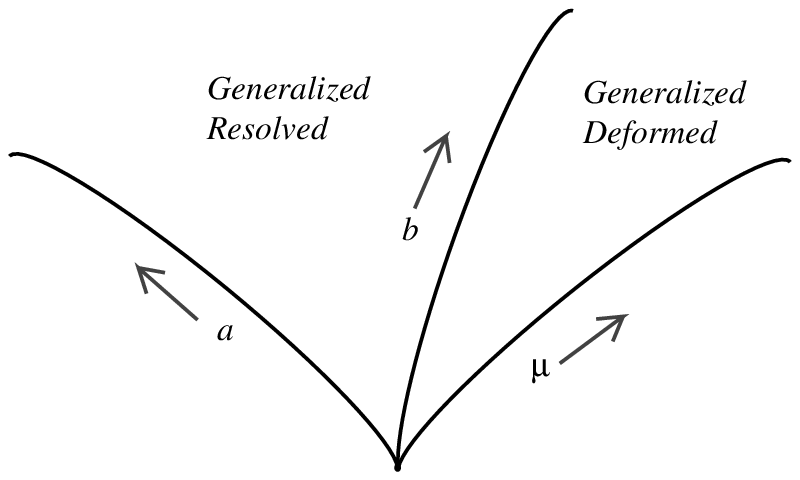}}
\vglue.1in
Fig. 1. Representation of the moduli space of generalized resolutions of
the conifold singularity. The two regions depicted correspond to the two
solutions of our constraint. The generalized deformed conifold metric is
singular. A point on each of the three lines represents, from left to
right, the resolved, regularized and deformed conifold. They meet at a
single point, the singular conifold.}}

\newsec{Summary and discussion}

In this letter, we have shown that lower dimensional gauged supergravities
are an appropriate framework to resolve singularities in the study of
gravity duals of supersymmetric gauge theories arising in D--branes that
wrap a supersymmetric cycle. The key ingredient is provided by the novel
twist prescription recently introduced in \eprii. The value of the gauge
fields at infinity implied by the conventional twisting is preserved,
the lower dimensional gauged supergravity governing the non-trivial
dynamics in the bulk. The singularity resolution takes place by switching 
on the appropriate fields of the gauged supergravity which correspond  to
the generalized twisting. 

We have presented a unified scenario for conifold singularity resolutions
from the perspective of M--theory: a single system encompassing both the
generalized resolution and deformation of the conifold singularity, each
of them emerging as the only two possible solutions of an algebraic
constraint. It might be possible to relate this constraint to those
appearing in the study of $G_2$ manifolds carried out in \eprii\ by a
reduction mechanism of the sort discussed in \hsiii.

It would be interesting to understand the meaning of $b$ on the
dual five dimensional gauge theory. In the regularized conifold it
plays the r\^ole of a mass scale: If a stack of D3--branes and
fractional branes is at the tip of the conifold \ks, $b \neq 0$
breaks the otherwise conformal invariance associated to the AdS
factor for small $\rho$. See, for example, \luvp. 

The mechanism presented in this letter must be useful in studying other
singularity resolutions. It would be also interesting to understand the
appearance of cascading solutions with chiral symmetry breaking occuring
in the IR --through the resolution of naked singularities-- \ks, in the
framework of gauged supergravity. We hope to report on some of these
issues in a near future.

\medskip
\bigskip
\noindent
{\bf Acknowledgements:}

It is a pleasure for us to acknowledge valuable comments from
Rafael Hernandez, Carlos N\'u\~nez, Leopoldo Pando Zayas and 
Konstadinos Sfetsos.
This work has been supported in part by MCyT and FEDER under grant
BFM2002-03881, by Xunta de Galicia, by Fundaci\'on Antorchas and
by Fundac\~ao para a Ci\^encia e a Tecnologia under grants
POCTI/1999/MAT/33943 and SFRH/BPD/7185/2001.

\listrefs

\bye

\listrefs

\end